# The Principle of Stationary Action in Biophysics:

# Stability in Protein Folding

## Walter Simmons

Department of Physics and Astronomy

University of Hawaii at Manoa

Honolulu, HI 96822

## Joel L. Weiner

Department of Mathematics

University of Hawaii at Manoa

Honolulu, HI 96822

October, 2013

# ABSTRACT

Processes that proceed reliably from a variety of initial conditions to a unique final form, regardless of moderately changing conditions, are of obvious importance in biophysics. Protein folding is a case in point. We show that the action principle can be applied directly to study the stability of biological processes.

The action principle in classical physics starts with the first variation of the action and leads immediately to the equations of motion. The second variation of the action leads in a natural way to powerful theorems that provide quantitative treatment of stability and focusing and also explain how some very complex processes can behave as though some seemingly important forces drop out.

We first apply these ideas to the non-equilibrium states involved in two-state folding. We treat torsional waves and use the action principle to talk about critical points in the dynamics.

For some proteins the theory resembles transition state theory (TST). We reach several quantitative and qualitative conclusions.

Besides giving an explanation of why TST often works in folding, we find that the apparent smoothness of the energy funnel is a natural consequence of the putative critical points in the dynamics. These ideas also explain why biological proteins fold to unique states and random polymers do not.

The insensitivity to perturbations which follows from the presence of critical points explains how folding to a unique shape occurs in the presence of dilute denaturing agents in spite of the fact that those agents disrupt the folded structure of the native state.

This paper contributes to the theoretical armamentarium by directing attention to the logical progression from first physical principles to the stability theorems related to catastrophe theory as applied to folding. This can potentially have the same success in biophysics as it has enjoyed in optics.

Descriptive Introduction

Processes that proceed reliably from a variety of initial conditions to a unique final state, regardless of changing conditions, are of obvious importance in biophysics. Proteins in an appropriate solution fold to unique forms and serve as a flagship example of stable processes in biology.

In this paper, we suggest how the action principle in classical mechanics could be used to analyze the stability of the protein folding process, which is of obvious importance per-se, but because the techniques described here follow from fundamental physics this approach will also be useful in the study of the stability of other biophysical processes.

In this introduction, we present a number of technical issues in a descriptive style. Technical details are discussed in a later section.

The action principle is a traditional starting point for classical mechanics. The action is a path integral of the difference between kinetic and potential energy, (the Lagrangian), between an initial and final time over a trajectory $S(t) = \int_{t_0}^{t} (T - V) dt$ . (The trajectory is implicit here.) The action is a scalar. The energy terms are written in generalized coordinates which take into account some or all constraints on the motion. The use of generalized coordinates makes this formalism particularly suited to moving parts of a complicated mechanical system. Standard treatments of the action principle allow for time-dependent potentials, which is also convenient for complicated processes.

Direct applications of the action principle, (i.e. without necessarily using the equations of motion that arise when the first variation of the action is set equal to zero), usually entail successive approximations [1].

When applied to mechanics the vanishing of the first variation of the action immediately yields the (Newtonian) equations of motion. The physical picture thus described is that of particles moving along trajectories according to the equations of motion.

The most important degrees of freedom in protein molecules are dihedral angles associated in pairs with amino-acid residues. In a common protein there might be 500 or more such angles. In folding, the molecule starts in some random assortment of these angles and moves toward a specific native set of angles. We speak of this motion as taking place in the space of dihedral angles.

If one considers protein molecules moving along trajectories in dihedral angle space, then several things are clearly missing from the trajectory picture.

First, the trajectories evidently move toward the common end point or points along the way to the native state but there is nothing explicit in the trajectory itself to define such a convergence; an energy landscape [2] , [3], [4], [5] is usually invoked to funnel the trajectories toward the end state.

Second, trajectories coursing through a rough energy landscape would arrive at the end point over a range of times; that is, diffusion. In contrast, many molecules have narrow melting curves and fast folding times that seem more appropriate to gas phase chemistry (TST). This is currently approached by postulating that the energy landscape [3], [6] is sufficiently smooth.

Third, there is the stability of the folding process. Consider an unfolded molecule in a dilute solution of a suitable denaturing agent. Such an agent interferes with the stability of the native state of the molecule but, curiously, does not deflect the process into alternative folded forms. Similarly, many other perturbations have little or no impact upon the final folded form.

Fourth, there is the problem of initial conditions. In the current view, a folding-ensemble of denatured protein molecules begins at the top of a funnel shaped energy landscape and proceeds down the funnel to the unique native state. The various conformations at the top of the funnel are equivalent in the sense that setting various initial conditions or subjecting the molecules to various perturbations results in conformations that are still in the folding-ensemble. The trajectory picture, per-se, does not address ensemble behavior; again,[7], [8], the energy landscape is invoked to explain how all the molecular trajectories behave in the same way.

These issues can all be addressed from a fundamental physics starting point by considering the vanishing of the second variation of the action,[9], [10], [11], [12]. This is an approach which has had spectacular success in modern optics wherein light rays focus, especially to a caustic [13].

Before we proceed we need to define some terms. Recall that the action is function on arcs. If the first variation with fixed endpoints is zero, then we call that arc a critical arc. If the second variation is positive for that arc, then that arc is a local minimum of the action. It is often convenient to regard the arc as part of a longer trajectory.   If we fix one endpoint of the arc at a point of the trajectory and move the other along the trajectory away from the fixed endpoint we may reach a point, and thus determine an arc, for which the second variation is zero.  If we move the movable endpoint even further away from the fixed endpoint the second variation will become indefinite, i.e. it can take on both positive and negative values.  Typically when we have a family of trajectories starting from a fixed point or fixed initial curve or surface, they will form a envelope, that is a curve or surface to which all the trajectories are tangent, and the points of tangency will be  points along the trajectories at which  the second variation vanishes.  This envelope is referred to a ``caustic." If the envelope happens to be a point, then we call that point a ``focus." When we have such an envelope, it dominates the motion in the sense that all of the trajectories meet it or pass through it.  (In a later section of this paper we cover this subject again in more mathematical detail.)

Some excellent examples of caustics in classical mechanical systems can be found in [9]

The concept of convergence is not, as we have just said, contained explicitly in individual trajectories.  Rather, the concept of convergence or focusing of mechanical trajectories is best described by considering families of trajectories.  If the dynamics of particles entails a caustic, then it is possible in principle to understand how a family of trajectories can behave in a coherent manner.

We next proceed to explain how powerful theorems of R. Thom and V. Arnold can be used to understand this behavior quantitatively.

We shall not attempt to define the stability of a shape in this paper. (We refer the reader to the literature section, below.) However, for this discussion of protein folding, stability means that the topology of a part of the native state (or of an intermediate state) is not altered by perturbations. See [14] for a similar concept.

We need two additional technical terms to be used in describing the action: state variable and control parameter. We do not require the mathematical definitions of these terms, but those definitions are readily available in the literature.

A simple way to look at state variable in a mechanical system is to think of space and time coordinates that are used to describe the motion. At a point in space and time, the description of the physical system will depend upon various control parameters. These may describe the interface with some apparatus, for example. For our purposes, the control parameters in folding are not tightly defined. The shape of the caustic will be defined entirely in terms of control parameters. They are assumed to be constant after folding and may turn out to be measurable distances or angles in the native state. Excellent examples of state variables and control parameters can be found in the literature, e.g. [15].

The mathematics tells us that, under appropriate constraints, there is a finite set of stable forms of the action near a critical point. Natural selection has evidently picked out these stable forms for biological molecules by choosing dynamics containing critical points. The stability arises because an ensemble of actions can change into one another as a result of a perturbation but the topology is not affected.

A simple example of this is the familiar cusp,

$$S(s) = \frac{1}{4}s^4 + \frac{1}{2}as^2 + bs \qquad (1.1)$$

Consider the case where this is topologically stable and where $s$ is a state variable and $a$, and $b$ are some variables (control parameters) that appear in the Lagrangian. Then the remarkable thing about this form is that for fixed values of $a$ and $b$ all possible perturbations have already been accounted for [11], [15]. So, if this is the action around the point $s=0$ then perturbations that have the form of higher order polynomials in $s$ do not change the shape of $S(s)$. This highly non-obvious result means that a trajectory or trajectories passing through a moderately rough energy landscape will not be topologically perturbed by small changes in the terms in the state variables (other than the two included in equation (1.1)). The other remarkable fact is that this cusp is one of only seven polynomial forms which have this remarkable property. If we suppose that folding occurs only when the action takes a multi-trajectory form, (and natural selection has eliminated unstable folds), then there are only seven distinct types of critical points. The possible trajectories fall into families or ensembles that have the property that perturbing one member of the ensemble changes it into another member of the same ensemble.

Finally, we note that the physics and mathematics show that when critical points are present in a dynamic, the critical point dominates the motion.

To summarize what have described so far:

1.) We start with the principle of stationary action applied to the dynamics of protein molecules.

2.) To account for folding we turn to a standard formalism for focusing.

3.) Two types of focus appear in the formalism: stable and unstable. We assume that natural selection has eliminated unstable foci.

4.) Thom's theorems now tell us that there are just seven possible functional forms for the action at a focus. Thom also tells us that these actions are stable against perturbations

Said differently, we are shifting our attention away from individual trajectories in dihedral angle space, with particles propagating according to the equations of motion, and toward groups of trajectories that share a common multi-trajectory action and which converge in dihedral angle space.

This completes a descriptive introduction to the idea of a critical point in the molecular dynamics. Before continuing this subject in more detail, we next discuss the folding process that is under examination in this paper.

## Two-State Folder and Torsion Waves

In this section, we set up the folding problem that we wish to address in a subsequent section.

We shall focus our analysis upon two-state folders; in particular, we are interested in the non-equilibrium transitions between the denatured and folded or intermediate states. [16], [17]

The molecules are not under over-all tension, so transverse waves and resonances with wavelength comparable to the length of the chain are disfavored. Torsional motions, which might include some long range waves, are favored by the geometry. A plausible picture is that energy is released at various localized points resulting in waves of torsional contraction or expansion which propagate away from the production point, generally with attenuation.

The theory described here does not depend upon the torsional form of the waves.

The details will ultimately depend upon whether the waves scatter off one another. In an earlier work on a continuous backbone model, the present authors showed that solitons are a possibility. Solitons pass through one another without shape change [18].

As we have said, the action is a scalar which depends upon energy and upon the path taken by a particle. For the two state folders, the action will depend upon the path taken by a molecule from unfolded to folded states. This path may be thought of as occurring in dihedral angle space. The molecule starts with a set of dihedral angles. It changes conformation following a path through dihedral angle space for which the first variation of the action is zero.

Toy Model

At this juncture, we pause to introduce a toy model which is solely intended to illustrate our points (and not to address the hard realities of folding dynamics [19]).

Let us simplify the torsional wave motion to just one axial degree of rotational freedom; i.e. an angle, $\theta$, describing a torsional shear, which will serve as an overly simplified generalized coordinate and it serves to allow us to construct a model action.

It is common in simulations of folding to introduce angular spring potentials $V(\theta) = k\theta^2$ for dihedral angles; these potentials depend upon the sequence. Critical points appear where we have a multi-trajectory action, as we have emphasized, above. To introduce that, we make the spring force asymmetrical. The force needed to turn a given dihedral angle depends upon the direction of turning and the angle at which it sits.

If we place the critical point, $\theta = 0$ at the folded end of the dynamical path then the kinetic energy at that end point is negligible.

Fortunately, for our purposes, a static version of this mechanical arrangement is well known in the catastrophe theory literature, where it is known as the Zeeman machine [15]. If natural selection rejects all unstable folding motions (and if this rotation is important in folding) then it turns out that the cusp in equation (1.1) with $\theta = s$, describes the potential energy, including any additional perturbations. Setting the kinetic energy aside for illustration we can now construct an action.

$$S(\theta) = \frac{1}{4}\theta^4 + \frac{1}{2}a\theta^2 + b \qquad (1.2)$$

In this case, $a$ and $b$ will be sequence dependent. The exact relationship between those parameters and the spring constants and lever arms are found in [15]. Introducing mutations to a given molecule could either change $a$ and/or $b$ or disrupt the form (1.2) altogether.

Obviously, at $\theta = 0$, $\delta S = \delta^2 S = 0$ as per our hypothesis.

If this were a valid theory of the torsional response to a wave passing through, then that response would be independent of modest perturbations other than the last two terms.

We could also use this potential to construct probability distributions and to derive statistical moments such as $\langle \Delta \theta^2 \rangle$. The moments derived from (1.1) are generally of simple form and change with $a$ and $b$. Note, of course, we have not specified the action for the entire molecule here. The $\varphi$ analysis of mutations will depend upon other parts of the molecule as well as this short segment. The details of moment analysis for various catastrophes are worked out in detail in [11].

We emphasize that this is a toy model which illustrates how a wave on a molecule can develop a critical point and be used in some calculations of measurable quantities and shows how the shape can be independent of perturbations, perhaps such as dissipative forces, energy rough spots, etc.

The toy model has no detailed structure (i.e. no sequence). However, it has unsymmetrical forces that can give rise to critical points and thereby, to stability. Note that the spring forces that are often used in simulations do not have these properties.

Addressing the Issues

With our descriptive introduction complete, we can now address the four issues listed above. We start with the assumptions that there is a critical point in the molecular dynamics and that natural selection has picked out stable folds.

The first point, that trajectories converge, is a direct implication of the presence of the critical point.

The explanation of the remaining three issues, (that the energy landscape is apparently smooth, that the folding process is stable under modest perturbations, and that the initial conditions in the denatured state do not matter very much), follow from the insensitivity of the action to perturbations. The energy landscape may have many rough spots but if they are not too extreme, then they do not change the multi-trajectory action and hence do not change the time to reach the folded state.

The time to the folded state (or to an intermediate state) for a short segment of the protein is a result of two important factors: (i.) a Boltzmann factor describing the escape from a potential well into the transition state, (ii.) a microscopic local rate factor, $\gamma$, describing how long it takes atoms moving on a fixed trajectory to collide and bind. The rate in the unfolded to folded direction takes the form,

$$Rate \to \gamma \exp(-\frac{\Delta G^{\dagger}}{k_B T}) \qquad (1.3)$$

where $\Delta G\dagger$ is the transition state free energy.

This picture of motions that occur along multiple similar trajectories until contact and bond formation is compatible with the observations that the time to folding is roughly proportional to contact order [20], [16],[21]. The factor $\gamma$ in equation (1.3) increases with distance along the chain and hence with contact order.

As can be seen directly from equation (1.3), a linear dependence of the free energy upon the concentration of denaturant might look like $\Delta G^\dagger = (\Delta G_0^\dagger)C$ giving chevron plots. It is also apparent that some proteins can fold at very high rates since the atoms only have to move along specific trajectories allowed by the presence of the critical point. Said differently, diffusion takes place before the molecules cross the barrier but follow trajectories toward a specific point (in action space) thereafter. Of course, not all proteins fit this simple picture.

Analysis Continued

An observation that follows the semi-quantitative description that we have presented so far is that some simplifications in folding result from the presence of a critical point in the molecular dynamics. For example, for two-state folders, the denatured and folded forms can both exist in equilibrium. The denaturing agent may impact the entropy but not the degrees of freedom associated with the folding. (This phenomenon is more general than protein folding. Catalysts that change the rate of a reaction by many orders of magnitude, by changing the heat flow to the thermal reservoir, without changing the reaction products are well known [22].)

Torsion waves on molecules in solution are expected to dissipate energy. The reliability of folding in the presence of agents that change the entropy or viscosity suggests that the degrees of freedom that participate in folding in an essential way are not impacted by dissipation. The theory presented here explains that using a combination of critical points and natural selection.

Another application of our theory is to address the question of why biological proteins fold to unique final forms while random polymers do not. Our theory suggests qualitatively that the former have critical points in the dynamics and fold along specific sets of trajectories while the latter do not have critical points and fold diffusively to various end shapes. Another way look at this is that the energy landscape may be rough for random polymers and they do not share the immunity to perturbations of biological proteins.

The topomer-sampling theory of Debe and collaborators, [23], considers folding in the restricted space of topomers (smooth transformations of the native conformation). That limits the number of degrees of freedom needed for diffusive search. (We remark that Thom's theory of generic stability uses functional forms that are interrelated through smooth coordinate changes (diffeomorphisms)). Our theory makes a stronger statement than topomer-sampling theory which is, that the action of the paths have critical points; allowable action must support multiple trajectories.

There are several alternative explanations for why trajectories that pass through a caustic continue on to the native state. One is that the caustic is small and it sits in a steep part of the energy funnel not far from the minimum. A similar phenomenon is the formation of an alpha helix. There is an initial energy barrier, but one that is passed, the helix quickly falls into place.

Our theory has not reached the point in development where the sequence dependence can be pinned down, nor have we identified the dynamical relationship between critical points and specific folds. However, some comments are in order. For torsion waves, this theory clearly requires that the sequence influence the mechanical parameters of torsional motion. An important feature is multi-trajectory action. A single trajectory theory, like TST for gas phase reactions, will not develop critical points; critical points are of essence due to multiple alternative paths.

A major difference between our theory and others is that here the vanishing of the second variation of the action is utilized to make connections to the existence of envelopes, i.e. caustics, and hence to catastrophe theory.

Physics, Mathematics, and Literature

This section is a concise treatment of the physics and the mathematics. We document this with references, especially books, where appropriate.

General references are as follows:

> For the mathematics is [10].
>
> For catastrophe theory is [11]. Other useful treatments of catastrophe theory are [15], [24], and, for catastrophe theory in chemical kinetics, [25].
>
> For the physics are [13], [12], and [9]. The physics and mathematics explained in the context of optics is found in [13].
>
> For completeness we mention that the action principle, including the formation of caustics, can be derived as a short wavelength limit of quantum mechanics [15], [11]. This is explicit in the book by Schulman, [26] wherein the Feynman path integral formulation of quantum mechanics is used, (especially Chapter 15 on caustics in quantum mechanics). We do not use quantum mechanics in this paper.

For protein science we suggest [22] and [27]. Additionally, various review papers some of which are cited in the text, e.g. [5], [19], [3], [28], [29], [30].

Early ideas underlying this work can be found in [31].

A particularly useful application of the calculus of variations to mechanics most commonly goes under the name of Hamilton's principle of least action, or stationary action. Other comparable principles exist but will not be discussed here [9]. In simple situations, Hamilton's principle can be stated as follows: Among all possible trajectories which take a single particle from a fixed initial position at a fixed initial time $t_0$ to another position at moment $t_1$, the realized motion is that for which the action integral

$$S = \int_{t_0}^{t_1} L(x, \dot{x}, t) dt \qquad (1.4)$$

is stationary. The integrand $L(x,\dot{x},t) = \frac{1}{2} m\dot{x}^2 - V(x,t)$ is referred to as the Lagrangian and the variable $x$ may be a generalized coordinate.

In order to apply Hamilton's principle we need to introduce variations of a particular arc that will be denoted by $x_0(t)$. We do this by introducing an arbitrary function $\xi(t)$, with $\xi(t_0) = \xi(t_1) = 0$, a real parameter $\varepsilon$ which is usually viewed as being very small, and considering the family of curves

$$x_\varepsilon(t) = x_0(t) + \varepsilon \xi(t) \qquad (1.5)$$

One thus obtains the action as a function of $\varepsilon$, $S(\varepsilon)$. The implication of the first variation being stationary, i.e. $\frac{dS}{d\varepsilon}=0$ for all possible families $x_\varepsilon$ is that the realized motions must satisfy what is called the Euler-Lagrange equation:

$$\frac{d}{dt}\frac{\partial L}{\partial \dot{x}}-\frac{\partial L}{\partial x}=0 \qquad (1.6)$$

We compute the second variation for curves that satisy the Euler-Lagrange equations, i.e. we compute $\frac{d^2S}{d\varepsilon^2}(0)$.

$$\frac{d^2S}{d\varepsilon^2}(0)=\int_{t_0}^{t_1}[-\xi^2\frac{\partial^2 V}{\partial x^2}+m\dot{\xi}^2]dt \qquad (1.7)$$

To deal with the integral one can expand $\xi(t)$ in terms of special functions appropriate to the behavior of $x_0$ and use their properties to simplify the above integral. We shall not explore that subject here.

When the second variation is positive, meaning that $\frac{d^2S}{d\varepsilon^2}(0)>0$, for all $x_\varepsilon$, one can show that $x_0$ minimizes $S$ compared to all arcs "close to" $x_0$ with the same endpoints.

To appreciate the sign of the second variation one wants to consider families of trajectories of realized motions. The simplest situation to consider is that for which the action $S$ is independent of the parameterization. We then consider such a family, where each trajectory begins at a particular point, or along a particular curve or on particular surface. Let's suppose all the motions are to begin along a curve which is parameterized using a parameter $u$. Then we can regard $S$ as a function of $u$ which identifies the initial point, and the position $x$, which is thought of as the terminal point of a curve realizing the motion; thus we write $S(u,x)$. To obtain an arc that achieves a minimum value of the action from the initial curve to a particular $x_1$, we solve $\frac{\partial S}{\partial u}(u,x_1)=0$ for $u$. If $u_1$ is a solution then $\frac{\partial S}{\partial u}(u_1,x)=0$ is the equation for an arc of a realized motion through $x_1$ that minimizes $S$ if $\frac{\partial^2 S}{\partial u^2}(u_1,x_1)>0$. If, on the other hand $\frac{\partial^2 S}{\partial u^2}(u_1,x_1)=0$, that is, the second variation is zero, this then identifies the presence of a caustic and $x_1$ is a member of the caustic.

The curves defined by $\frac{\partial S}{\partial u}(u,x)=0$ form an envelope and $x_1$ is the point at which the curve $\frac{\partial S}{\partial u}(u_1,x)=0$ through $x_1$ is tangent to the envelope. In higher dimensions, simple derivatives are replaced by partial derivatives and the second derivative is replaced by the Hessian, i.e., the matrix of second derivatives. The vanishing of the ordinary second derivative is replaced by the vanishing of the determinant of the Hessian.

In typical applications of singularity theory, or catastrophe theory, one considers a given function of several variables, which are referred to as state variables and control variables. One focuses on the form of sets determined by setting to zero the first and second derivatives of the given function with respect to state variables and takes advantage of known generic solutions of those equations for certain numbers of control variables. The function $S(u,x)$ introduced above, where $u$ is a state variable and $x$ is a control variable, fits this situation since we are clearly interested in sets where $\frac{\partial S}{\partial u}(u,x) = 0$ and $\frac{\partial^2 S}{\partial u^2}(u,x) = 0$. This is how one can establish a connection between calculus of variations and catastrophe theory.

Noticing that geometric optics fails in this situation, Berry and Upstill comment that this failure is 'catastrophic' because this is just the point at which catastrophe theory becomes applicable.

Where caustics are present we have a strong focusing of trajectories into a space that is very closely circumscribed by up to five control parameters [11]. A critical point, where the above matrix vanishes, dominates the dynamics in that neighborhood around the critical point.

The conditions on the partial derivatives just mentioned are the same as the conditions for catastrophe theory to obtain. There are many texts on catastrophe theory so we will just remark that stability of the catastrophe (caustic) against perturbations of the state variables (time and space) is the major result we have used.

Let us consider the limitations of this conjecture.

An important issue is the degree of sensitivity to perturbations. For example, the strength of the denaturing agents may be so great that our conjecture may not apply. There appears to be no general rule from catastrophe theory to quantify the limits of allowed perturbations; the answer is in the details

Another limitation is the possible appearance or non-appearance of false minima. Further research will be required to understand this issue. In a special case, however, if there is a single, long-lived false minimum, (with the molecules slowly leaking down to the thermodynamic minimum), then our conjecture may apply to the false minimum. We remark that prions may be such a case.

## General Comments

We emphasize that in this paper we are not addressing the issue of protein structure [22],[30]; rather, we are addressing the issue of the stability of the folding process, especially the earliest stage from denatured ensemble forward. Prospectively, a full understanding of the early phases is potentially very useful in experiment design where it is necessary to evaluate the impact of various external factors on the shape of the native state (e.g. Fluorescence resonance energy transfer (FRET), various denaturants, etc.), in the design or discovery of agents or environmental factors, [32], that interfere with folding, in engineering new proteins that fold in specific ways. Retrospectively, an understanding of the stability of the earliest phases of folding has potential value in the study of the last common ancestor and the origin of life.

## Future Directions

The calculations to check these ideas in model molecules are not simpler than traditional folding simulations. However, the results are different.

If the putative caustic appears somewhere along the folding path, then a simulation of the action up to that point can reveal it. The quantitative test is the vanishing (or near-vanishing) of the Hessian determinate, as described in the previous section. The number of computations is formidable but, for simple molecules, not beyond supercomputer capacity.

Should a caustic be indicated along the folding path, this can be confirmed by searching for a saddle point in the action as the integration is continued to a point just beyond the caustic. Again, the computations are formidable but not impossible.

Once a caustic is found in a model, it will become possible to tune model parameters to optimize folding.

## Conclusions

The phenomenon studied is the motion of protein molecules in a variety of initial conditions and, in the presence of various perturbations, terminating in a unique final state in spite of their being relatively little free energy available.

The principle of stationary action leads immediately to equations of motion. However, equations of motion describe the propagation of the molecule along trajectories in dihedral angle space and tend to obscure the behavior of groups of trajectories. Moreover, as noted originally by Levinthal, the number of conformations in dihedral angle space is of cosmological proportions.

By treating the problem directly using the principle of stationary action, and putting the equations of motion aside, it is possible to treat groups of trajectories that behave in a similar manner; in particular, trajectories that converge either to a focus or to a caustic.

The resulting treatments narrows the number of possible paths through dihedral angle space because all trajectories pass through very narrow caustics located somewhere along the folding path.

The result of direct analysis from the action is that two types of focusing emerge: stable and unstable. We assume that natural selection has eliminated the unstable focusing. This treatment leads immediately to the strong stability of the process of folding. Many features of the folding process emerge directly.

As in most biological processes, protein folding entails a large number of complicated forces and parameters that change with conformation (i.e. with time during folding) and, as just mentioned, it takes place in a space of very high dimension. Yet folding, does indeed lead to unique final forms even in the presence of denaturing agents that are chosen to disrupt the final shape.

This complexity might be enough to send any theorist back to his coffee pot. However, when this is approached from the viewpoint of the direct application of principle of stationary action, an idea takes shape rather naturally. The idea is that the fundamental dynamics of molecular motion contain critical points that dominate the motion and make the motion less vulnerable to disruption by various changing forces and conditions. This dominance of dynamical behavior by critical points is well established in physics.

We have shown how this idea emerges from the action principle and have given semi-quantitative explanations for many of the phenomena that have been documented in laboratories and in simulations over the past five or six decades.

For theories that start with the molecule in its native state, unfold it in the lab or in simulation, and then refold it, the acid test is prediction of the native form. By starting with the denatured state and applying physics and mathematics to study the stability of the folding process, we have only a germ of a full theory of folding and we cannot predict structures; not even approximately.

 No theories, which are consistent with classical mechanics, are in contradiction with least action and the vanishing of the first variation of the action for the dynamics of the molecule during folding. The fundamental departure embodied in this work is the putative vanishing of the second variation of the action, implying that various trajectories can be treated as a unit, and the role of natural selection in eliminating unstable folds.

What we have accomplished is to show that these putative critical points provide a level of quantitative understanding of many observed features: rapid rates (TST like behavior), smoothness of the energy landscape, non-folding of random polymers, insensitivity to many perturbations, and some qualitative insights in to other features of folding such as the importance of topology and contact order.

The obvious next steps in the development of this theory are to learn exactly how the critical points emerge in terms of sequence and to learn how the critical points relate to specific structures.

34